\documentclass[doublecol,figures]{epl2}
\usepackage[T1]{fontenc}
\usepackage[latin1]{inputenc}
\usepackage{graphicx}
\usepackage[english]{babel}

\title{Mechanisms of magnetoelectricity in manganese doped incipient ferroelectrics}
\shorttitle{Magnetoelectricity in Mn doped incipient ferroelectrics} 

\author{R.O.\ Kuzian \inst{1} 
   \and V.V.\ Laguta \inst{1,2}
   \and A.-M. Dar\'e \inst{3}
   \and I.V.\ Kondakova \inst{1}
   \and M.\ Marysko  \inst{2}
   \and L.\ Raymond \inst{3}
   \and E.P.~Garmash \inst{1}
   \and V.N.\ Pavlikov \inst{1}
   \and A.\ Tkach \inst{4}
   \and P.\ M.\ Vilarinho \inst{4}
   \and R.\ Hayn \inst{3}
}
\shortauthor{R.O.\ Kuzian \etal}
\institute{                    
  \inst{1} Institute for Problems of Materials Science NASU - Krzhizhanovskogo
     3, 03180 Kiev, Ukraine\\
  \inst{2} Institute of Physics, AS CR - Cukrovarnicka 10, 16253 Prague, Czech
     Republic\\
  \inst{3} Aix-Marseille Universit\'e, IM2NP CNRS - 
     FST St-J\'er\^ome, Avenue Escadrille Normandie Niemen, 
     F-13397, Marseille Cedex, France \\
  \inst{4} Department of Ceramics and Glass Engineering, CICECO, University
     of Aveiro - 3810-193 Aveiro, Portugal
}
\pacs{76.}{Magnetic resonance}

\abstract{
We report magnetization measurements and magnetic resonance data for
SrTiO$_{3}$ doped by manganese. We show that the recently reported coexistent
spin and dipole glass (multiglass) behaviours are strongly affected
by the distribution of Mn ions between the Sr and Ti sites. Motivated
by this finding we calculate the magnetic interactions between Mn impurities
of different kinds. Both LSDA+$U$ and many-body perturbation theory
evidence that magnetic and magnetoelectric interactions
are mediated by Mn$_B^{4+}$ ions substituting for Ti. We propose two
microscopic magnetoelectric coupling mechanisms, 
which can be involved in  all magnetoelectric systems based
on incipient ferroelectrics. In the first one, the electric field modifies
the spin susceptibility via spin-strain coupling of Mn$_{B}^{4+}$.
The second mechanism concerns Mn pairs coupled by the
position-dependent exchange interaction. 
}

\date{12.09.10}

\begin{document}

\maketitle

\section{Introduction}

SrTiO$_{3}$ (STO) and KTaO$_{3}$ (KTO) doped by manganese have attracted
considerable attention exhibiting simultaneous spin and dipole glass
behaviours with large non-linear magnetoelectric coupling 
\cite{Kleemann08,Kleemann09,Shvartsman08}.
Such "multiglass" systems extend non-trivially
the frame of conventional multiferroicity and give new perspective
for studies of the phenomenon and potential application in microelectronic
devices.

Both STO and KTO are special representatives of the perovskite family
of ABO$_{3}$ materials. They are incipient ferroelectrics (IF), i.e.
they remain paraelectric down to zero temperature, but exhibit very
large dielectric permittivity ($\sim$~20000 and 5000 respectively)
at low temperature due to the softening of a transverse optical mode
that corresponds to B sub-lattice oscillations with respect to the almost
static rest of the lattice.

The Mn impurities in STO may substitute both for Sr and Ti; they
will be denoted as Mn$_{A}$ and Mn$_{B}$ respectively. Isolated
impurities are paramagnetic, Mn$_{B}$ being an isotropic
centre with formal valency Mn$^{4+}$, and a spin $S=3/2$ \cite{Muller59},
while Mn$_{A}$ which has valency 2+ and $S=5/2$, is isotropic at $T>100$ K, and axial
at low temperature\cite{Laguta07}. According to the interpretation
of ESR measurements\cite{Muller59,Laguta07}, which were recently
confirmed by density functional theory (DFT) calculations
\cite{Kondakova09,Kvyatkovski09}
and EXAFS experiments\cite{Levin10}, the Mn$_{B}$ impurity resides
in the octahedrally coordinated cubic position B of the perovskite
lattice, and Mn$_{A}$ is displaced from A position 
thus forming electric dipoles in addition to magnetic ones.

Ceramic samples of STO doped by 2\%
of manganese exhibit spin- and polar- glass properties at temperature
below $T_{g}\approx$~38~K. Moreover, a substantial non-linear magnetoelectric
coupling was measured \cite{Kleemann08,Kleemann09,Shvartsman08}.
A similar behaviour for KTO:Mn system was also found \cite{Kleemann09}.
The interaction of electric dipoles formed by off-central Mn$_{A}$
impurities has the same nature as the interaction of other dipole
impurities in IF, its mechanism is rather well understood\cite{VuGlin90}.
In this paper we concentrate on magnetic and magnetoelectric interactions
in STO:Mn. We show that the presence of off-central Mn$_{A}^{2+}$
ions substituting for Sr is necessary but not sufficient to induce the multiglass
behaviour, and that the magnetic interactions are mediated by Mn$_{B}^{4+}$
ions substituting for Ti.

Some aspects of the considered problem are interesting from the fundamental point 
of view. The interaction between ions with different $d$-shell filling and different 
spins connected by several bridging ligands requires a generalization of
superexchange theory. The dependence of spin-Hamiltonian parameters on the external 
electric field is a non-trivial application of the ligand field theory for ions in a 
highly polarizable medium.

\section{Experiment}

For our experimental studies we have used two groups of STO:Mn ceramic
samples prepared in different labs. We will call them I and II. Type-I
ceramics with the formal chemical composition Sr$_{0.98}$Mn$_{0.02}$TiO$_{3}$
were prepared by mixed oxide technology described elsewhere\cite{Tkach05}.
In particular, reagent grade SrCO$_{3}$, TiO$_{2}$ and MnO$_{2}$
were mixed in appropriate amounts, ball milled, dried and calcined
at 1100~$^{\circ}$C for 2 h. The calcined powders were again milled,
pressed isostatically and sintered at 1500~$^{\circ}$C for 5 h. For the
type-II ceramics (formal chemical composition Sr$_{0.96}$Mn$_{0.04}$TiO$_{3}$),
instead of MnO$_{2}$, powder of MnCO$_{3}$ was used. The calcination
was performed at 1150~$^{\circ}$C for 4 h and the sintering at 1360~$^{\circ}$C
for 2 h. Both types of samples possess Mn$_{A}^{2+}$ off-central
impurities and have similar dielectric properties, but as 
shown below their magnetic responses are strikingly different.

\begin{figure}[htbp]
\onefigure[scale=1.0]{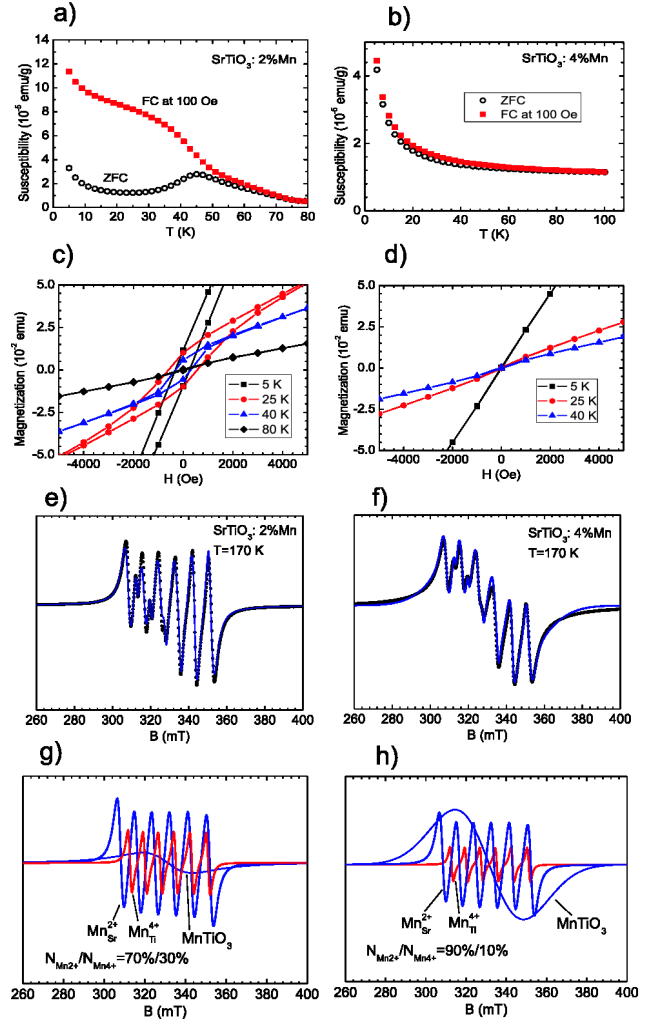} 
\caption{(Coloronline) Comparison of magnetic properties for ceramics I (left)
and II (right). a),b) ZFC and FC susceptibilities as a function of temperature.
Magnetization as a function of applied field, measured c) in type-I ceramics
at 80, 40, 25, and 5~K, and d) in type-II ceramics
at 40, 25, and 5~K. e),f) Experimental (dots) EPR spectra, and the corresponding
simulations (solid lines). g),h) EPR spectra decomposition which
shows the percentage of various Mn paramagnetic centres. \label{Fig1} }
\end{figure}

The magnetic measurements were performed using a SQUID magnetometre
(MPMS-6S Quantum Design) in the temperature range 4.5 - 100 K. The
zero-field-cooled (ZFC) and field-cooled (FC) susceptibilities were measured
in an applied field of 100 Oe. The main experimental result of this paper
is illustrated in Fig.\ref{Fig1} .%
While magnetic properties of type-I ceramics are similar to that reported
in Refs.~\cite{Kleemann08,Shvartsman08}, namely, they exhibit spin-glass
behaviour at $T<45$~K (Fig. 1a), the type-II ceramics remain paramagnetic
down to liquid helium temperature (Fig. 1b). In addition, for the
ceramics I the magnetization exhibits a hysteretic behaviour with
a finite remanence and coercivity (Fig. 1c). On the contrary, the
magnetic loops measured on ceramics II (Fig. 1d) show no hysteresis
and remanence. In this case, the magnetization is almost entirely 
determined by the paramagnetic contribution.

In order to study the distribution of Mn ions in the lattice and their 
individual magnetic properties -they are the only
source of magnetism in our system- we have performed ESR measurements
at 9.2 GHz in the standard 3 cm wavelength range at temperatures from
4.2 up to 300 K. An Oxford instrument ESR 900 cryosystem was used.
In both types of ceramics we have found the spectra of Mn$_{A}^{2+}$
and Mn$_{B}^{4+}$ ions, which were described in detail in our previous
paper and other early publications (see, e.g. Refs.~\cite{Muller59,Laguta07}).
The main difference between the two types of ceramics revealed by
ESR is the ratio of Mn$^{2+}$/Mn$^{4+}$ ion concentration, which 
is 70/30 and 90/10 for types I and II respectively, as shown in Fig. 1.e-h. 
The absolute concentration of Mn$_{A}^{2+}$ in both ceramics is about
1$\div$1.3 \% as the ceramics contain also 
some amount of MnTiO$_{3}$
inclusions reflected in ESR spectra by a broad line (see, Fig. 1 g,h)
and observed previously in electron diffraction spectra (Ref. \cite{Tkach05}).
Although in the type-II ceramics about 70\% of Mn ions
contribute to this MnTiO$_{3}$ fraction, 
it does not markedly influence the magnetic properties of the studied 
samples (Figs. 1b,1d) due to relatively small magnetic anomaly of MnTiO$_{3}$ 
at the antiferromagnetic phase transition at 63 K  (Ref. \cite{Stickler67}).

\section{DFT calculations}

We have theoretically studied
the interactions between Mn ions using LSDA+$U$ and many-body perturbation
theory. We have considered the Mn impurities when they occupy nearest
neighbour positions, and looked at the various pairs Mn$_{A}$-Mn$_{A}$, Mn$_{A}$-Mn$_{B}$,
and Mn$_{B}$-Mn$_{B}$.  The Mn$_{A}$ ions were
shifted in various directions from the symmetric A position.

Density functional theory calculations were performed using the
full- potential local-orbital (FPLO) code\cite{FPLO}. The total energies for different
configurations were obtained, and the results were mapped onto an Mn-Mn pair
effective Hamiltonian of the form 
\begin{equation}
\hat{H}=E_{n}\left(\mathbf{d}_{1},\mathbf{d}_{2}\right)+
J_{12}\left(\mathbf{d}_{1},\mathbf{d}_{2}\right)
\hat{\mathbf{S}}_{1}\hat{\mathbf{S}}_{2},
\label{eq:Heff}
\end{equation}
 where $\mathbf{d}_{i}$ is the shift from cubic position. $E_{n}$
is a non-magnetic spin-independent interaction, 
treating the ion motion classically,
it is
a $c$-number in our approach.
The spin part of the interaction is of Heisenberg-type.

We present the results for Mn$_{A}^{2+}$-Mn$_{A}^{2+}$ pairs in
Table~\ref{tab:I}. Various Mn$_{A}^{2+}$ displacements along
the symmetry axis $Z$ were considered: $d_{1}=d_{2}$ 
(the ions shift conserving the distance between them), 
$d_{1}=-d_{2} > 0$
(ions shift towards each other), $d_{1}= -d_{2} < 0$ (away from each
other). In each case the results correspond to the total energy minima
with respect to displacement. 
The non shifted distance between two A sites is 3.9 \AA\cite{Kondakova09}.
\begin{table}
\caption{\label{tab:I} Hamiltonian (1) parameters calculated with
LSDA+$U$ for various configurations of Mn$_{A}^{2+}$-Mn$_{A}^{2+}$
pair in a $2\times2\times3$ 60 atom supercell (symmetry group P4mm,
\#99); $U=$4~eV }
\begin{center}
\begin{tabular}{cccc}
\hline
$d_{1}$,\AA{}  & $d_{2}$,\AA{}  & $E_{n}$, meV  & $J_{AA}$, meV  \tabularnewline
\hline 
0  & 0  & 0  & -0.008  \tabularnewline
0.82  & 0.82  & -695.5  & 0.23  \tabularnewline
0.35  & -0.35  & -26.6  & 0.46  \tabularnewline
-0.47  & 0.47  & -27.2  & 0.04  \tabularnewline
\hline
\end{tabular}
\end{center}
\end{table}

The third column shows that the configuration with parallel
shift of Mn ions ($d_{1}=d_{2}$ = 0.82\AA) is separated from the others by
such a large energy, that these will not be observed
within the physically relevant temperature range. A similar situation
was described in Ref.\cite{Prosandeev03} for Li-Li pairs in KTaO$_{3}$.
The last column of Table \ref{tab:I} represents one of the main
theoretical findings of this work, namely, it shows that the magnetic
interaction of nearest neighbour Mn$_{A}^{2+}$ ions,
for the most probable parallel configuration,
satisfies $\left|J_{AA}/k_{B}\right|<3$~K.
In fact this value is
on the verge of precision of DFT calculations. Despite the large
spin value of Mn$_{A}^{2+}$ ions, this interaction cannot be responsible
for magnetic susceptibility anomalies at $T\sim40$~K that we observe
for type I ceramics (Fig.\ref{Fig1}), and that were reported in previous
studies of STO:Mn \cite{Shvartsman08,Kleemann08}.

Table \ref{tab:II} shows the results for a Mn$_{A}^{2+}$-Mn$_{B}^{4+}$
pair. Here the Mn$_{A}^{2+}$ ion shift was taken to be the same as
the one found for the isolated ion\cite{Kondakova09}; for $d_A >(<)0$ the ions get farther (closer).
\begin{table}
\caption{\label{tab:II}  Hamiltonian parameters for Mn$_{A}^{2+}$-Mn$_{B}^{4+}$
pair, calculated for a $2\times2\times2$ 40 atom supercell (P1, \#1);
$U=$4~eV }
\begin{center}
\begin{tabular}{cccc}
\hline 
$d_{A}$, \AA{}  & $E_{n}$, meV  & $J_{AB}$, meV  & $J_{AB}/k_{B}$, K\tabularnewline
\hline 
0  & 0  & 0.5  & 5.8\tabularnewline
-0.64  & -144.6  & 1.65  & 19.1\tabularnewline
0.64  & -123.02  & 0.51  & 5.9 \tabularnewline
\hline
\end{tabular}
\end{center}
\end{table}
Finally, for the pair Mn$_{B}^{4+}$-Mn$_{B}^{4+}$ we have found
$J_{BB}=9.3$~meV, i.e. $J_{BB}/ k_{B}=107.9$~K. So, we may conclude that the
presence of Mn$_{B}^{4+}$ions may be responsible for the observed anomalies
in magnetic susceptibilities.

\section{Theory of superexchange}
To reach a better understanding of the exchange mechanism
we have performed
analytic calculations of the superexchange interaction between Mn ions
within fourth-order many-body perturbation theory. Using resolvent
method \cite{Auerbach} we obtained a general formula for  the magnetic
coupling between two spins $S_{1}$ and $S_{2}$ in terms of
hopping integrals $t_{im\beta n}$ between cations $(im)$ ($i$ specifies
the cation, $m$ the orbital index) and neighbouring ligands $(\beta n)$
($\beta$ specifies the ligand, $n$ the orbital). Taking the hole
point of view (no fermion on the ligand $p$ orbitals in the ground
state) we have established Eq.(\ref{Jab}),
\begin{widetext} 
\begin{eqnarray}
J & = & \frac{1}{2S_{1}S_{2}}\sum_{\beta\beta^{\prime}nn^{\prime}}
\sum_{mm^{\prime}}^{occ}t_{1m\beta n}t_{2m^{\prime}\beta n}
t_{1m\beta^{\prime}n^{\prime}}t_{2m^{\prime}\beta^{\prime}n^{\prime}}
\left\{\frac{1}{\Delta_{1\beta n}\Delta_{1\beta^{\prime}n^{\prime}}\Delta_{12}}
+\frac{1}{\Delta_{2\beta n}\Delta_{2\beta^{\prime}n^{\prime}}\Delta_{21}}
\right.\nonumber \\
 & + & \left.\frac{1}{\Delta_{1\beta n}+\Delta_{2\beta^{\prime}n^{\prime}}+
 U_{p}\delta_{\beta\beta^{\prime}}}\left(\frac{1}{\Delta_{1\beta n}}+
 \frac{1}{\Delta_{2\beta^{\prime}n^{\prime}}}\right)\left(\frac{1}{\Delta_{2\beta n}}
 +\frac{1}{\Delta_{1\beta^{\prime}n^{\prime}}}\right)\right\}, \label{Jab}
 \end{eqnarray}
\end{widetext} 
where the sum over the cation orbitals $m$ and $m^{\prime}$
is restricted to ground-state occupied ones (abbreviation "occ"). $\Delta_{i\beta n}$
is the energy of the excited state (measured with respect to the ground
state), where one fermion has moved from $(im)$ to $(\beta n)$,
while $\Delta_{ij}$ is the difference of energy between an excited
state with $N_{i}-1$ and $N_{j}+1$ fermions on cation $i$ and $j$
respectively, and the ground state (GS) with $N_{i}$ and $N_{j}$ fermions
per cation respectively. If the cations are of the same type, one has $\Delta_{ij}=\Delta_{ji}$
and $\Delta_{i\beta n}$ does not depend on $i$. The three added
terms inside the bracket in Eq.(\ref{Jab}) correspond to different
paths: the first two involve intermediate excited states with $(N_{i}-1,N_{j}+1)$
fermions on the cations, while the third term corresponds to an excited
state with two fermions on the ligands with repulsion $U_{p}$ when
the holes meet on the same ligand.

The parameters $t_{im\beta n}$,  $\Delta_{i\beta n}$ and $\Delta_{ij}$
were extracted from the analysis of photoemission experiments reported in 
Refs.\cite{Bocquet92g,Mizokawa48g}.
We denote for the BB-pair: $\Delta_{12}=U_{eff}^{B}$, 
$\Delta_{i\beta n}=\Delta_{eff}^{B}$;
for the AA-pair $\Delta_{12}=U_{eff}^{A}$, $\Delta_{i\beta n}=\Delta_{eff}^{A}$;
while for the BA-pair $\Delta_{12}=\Delta_{eff}^{B}-\Delta_{eff}^{A}+U_{eff}^{A}$,
$\Delta_{21}=\Delta_{eff}^{A}-\Delta_{eff}^{B}+U_{eff}^{B}$, 
$\Delta_{1\beta n}=\Delta_{eff}^{B}$,
and $\Delta_{2\beta n}=\Delta_{eff}^{A}$, where 
$\Delta_{eff}^{i}=\Delta_{i}+J_{Hi}\left[7/9(N_{i}-1)-p_{i}\right]$
and $U_{eff}^{i}=U_{i}+J_{Hi}\left[N_{i}+11/9-2p_{i}\right]$,
with $J_{Hi}=5/2B_{i}+C_{i}$, $N_i$ is the number of holes in the ground
state of Mn$_{i}$ ion and $p_{i}$ is the number of doubly-occupied
orbitals. We use the following experimental values (in eV) for Mn$_{B}$:
$B_{i}=0.132$, $C_{i}=0.610$, $\Delta_{i}=2\pm0.5$ and $U_{i}=7.5\pm0.5$
(Table I of  Ref.\cite{Bocquet92g},), while for Mn$_{A}$, $B_{i}=0.119$,
$C_{i}=0.412$, $\Delta_{i}=7$ and $U_{i}=5.5$ (Table I, II of 
Ref. \cite{Mizokawa48g}). This gives $\Delta_{eff}^{A}=9.2$eV, $\Delta_{eff}^{B}=4.5\pm0.5$eV,
$U_{eff}^{A}=9.9$eV and $U_{eff}^{B}=11.6\pm0.5$eV. The hopping
parameters are expressed in terms of Slater-Koster parameters \cite{Harrison}
$V_{pd\pi,\sigma}^{A(B)}$ for Mn$_{A(B)}$, they essentially depend
on the cation-ligand distance, we take $V_{pd\sigma}^{i}/V_{pd\pi}^{i}=-2.16$
\cite{Harrison}. Following Ref.\cite{Mizokawa48g},
we have assumed that the hopping which links states with $N+1$ and
$N$ holes on some cation, is reduced by a factor $R$ compared to hopping
linking states with $N-1$ and $N$ holes. The results 
(and their comparison with LSDA+$U$) are presented in Table \ref{tab:III}
for the parameters $R=1.2$ and $U_{p}=4$~eV, and the shifts are the same as those found in LSDA calculations. %
\begin{table}
\caption{\label{tab:III} Superexchange values $j_{BA}= J_{BA}/J_{BB}$, 
and $j_{AA}=J_{AA}/J_{BB}$
with $J_{BB}=(9.5 \pm 1.5)(V_{pd\pi}^{B})^{4}$ for different A-shift.
}

\begin{center}
\begin{tabular}{ccccc}
\hline 
$d_{A}$   
& $j_{BA}$& $j_{BA}$,& 
$j_{AA}$ & $j_{AA}$, \tabularnewline
& & LSDA+$U$   & & LSDA+$U$  \tabularnewline
\hline 
0       &  0.39 & 0.054 & 0.008 & -0.001\tabularnewline
0.82    & \ & \ & 0.008 &  0.025 \tabularnewline
-0.64   & 0.52 &  0.177  & \ & \ \tabularnewline
0.64    & 0.29 & 0.055 & \ \tabularnewline
\hline
\end{tabular} 
\end{center}
\end{table}
 
As can be seen,  the agreement is qualitative between perturbative calculations
and DFT results. In addition, the super-exchange theory enables  to
discuss various tendencies and contributions. The different exchange
coupling $J_{BB}$, $J_{AB}$ and $J_{AA}$ are proportional to $(V_{pd\pi}^{B})^{4}$,
$(V_{pd\pi}^{A})^{2}(V_{pd\pi}^{B})^{2}$ and $(V_{pd\pi}^{A})^{4}$
respectively. 
A crude estimate based on the different distances, leads to  $J_{AB}/J_{BB}\sim0.1$
and $J_{AA}/J_{BB}\sim0.01$. 
One can also expect a predominant value for
$J_{BB}$ due to  geometry: it corresponds to a 180$^{\circ}$ Mn$_{B}$-O-Mn$_{B}$ link,
while the other two 
are 90$^{\circ}$ ones. 
The difference between exchange couplings are also due to different number of exchange paths
and to difference in charge transfer values $\Delta_{eff}^{A} > \Delta_{eff}^{B}$.
The number of paths for $J_{AA}$ is larger than for the other two, since
four ligands and many orbitals come in, against three ligands for $J_{AB}$
and only one for $J_{BB}$.
Two or more ligands raise the opportunity to have
ferromagnetic contributions, which explains reduction of $J_{AA}$
compared to the other two, further than what was expected just
by distance effect.

\section{Mechanisms of magnetoelectric coupling}

We propose two possible mechanisms involving Mn$_{B}^{4+}$
ions, which may be relevant for IF doped by manganese.
In the following
we consider the solid solution Sr$_{1-x}$Mn$_{x}$Ti$_{1-y}$Mn$_{y}$O$_{3}$.
The magnetoelectricity implies the dependence of the magnetic susceptibility on an 
electric field. 
Up to second order
, the magnetic susceptibility may
be written 
\begin{equation}
\chi_{ij} = -\frac{\partial^{2}F}{\partial H_{i}\partial H_{j}}\biggr|_{H=0}
=\chi_{0,ij}+ \chi_{1,ijk}E_{k}+\chi_{2,ijkl}E_{k}E_{l}.\label{eq:chiE}
\end{equation}
where the free energy density $F$, the magnetic and electric field
components, $H_{i}$ and $E_{i}$ are measured in CGS units.
To translate this equation in terms of 
Ref.\cite{Shvartsman08} notations with SI units,
one should use the relations 
\begin{eqnarray}
\beta_{ijk} & = & \left(4\pi\right)^{\frac{3}{2}}\mu_{0}
\sqrt{\varepsilon_{0}}\chi_{1,ijk}
\approx1.67\cdot10^{-10}\chi_{1,ijk}\:\mathrm{s/A},\label{eq:chi2bet}\\
\delta_{ijkl} & = & 8\pi^{2}\mu_{0}\varepsilon_{0}\chi_{2,ijkl}
\approx8.78\cdot10^{-16}\chi_{2,ijkl}\:\mathrm{sm/VA}.\label{eq:chi2del}
\end{eqnarray}

The first mechanism
 is a one-spin effect:
The polarization $P$ of the lattice,
is accompanied by a lattice strain, which is proportional to the
square of polarization\cite{Rimai62}.
The Mn$_{B}^{4+}$ ion has a $3d^{3}$ configuration and a $^{4}F$ GS, 
which is split by the ligand field in contrast to the $3d^{5}$ 
configuration and $^{6}S$ GS
for Mn$_{A}^{2+}$. In a cubic field the Mn$_{B}^{4+}$ ion has a fourfold degenerate $^{4}A_{2}$ GS. 
When the local symmetry becomes axial, an additional splitting 
arises. Its magnitude depends on the polarization via the strain. This affects 
the magnetic susceptibility. 
In a paraelectric phase, the changes are proportional to the 
square of the external electric field, but in the presence of a net polarization 
$P_r$ in a polar
phase, a linear dependence appears. 
Let us note that this mechanism will be effective in any ferroelectric perovskite 
doped by paramagnetic ions located at B sites and having the $3d^3$ configuration.

The contribution of this mechanism to the magnetoelectric susceptibility is given 
below by Eqs.(\ref{eq:chi1app}) and (\ref{eq:chi2zzzz}). Substitution of numerical
values relevant to the experimental conditions of Ref.\cite{Shvartsman08}
gives 
\begin{eqnarray}
\chi_{1,zzz} & \approx & -2.2\cdot10^{-9},\;
\beta\approx-3.6\cdot10^{-19}\:\mathrm{s/A},\label{eq:chi1val1}\\
\chi_{2,zzzz} & \approx & -1.2\cdot10^{-10},\;
\delta\approx-0.1\cdot10^{-24}\:\mathrm{sm/VA}.\label{eq:chi2val1}
\end{eqnarray}
In these estimates,
 we have used $x+y=0.02$, and $x/y\approx70/30,$
as found in our ESR experiment. 
The reported
experimental values are\cite{Shvartsman08} $\beta\approx-3\cdot10^{-19}\:\mathrm{s/A}$ and
$\delta\approx-9\cdot10^{-24}\:\mathrm{sm/VA}$. We see that the  paramagnetoelectric
susceptibility $\beta$ is in fairly good agreement, but the biquadratic
coefficient is underestimated by this mechanism which cannot be the only argument put forward.

The second mechanism concerns Mn$_A$-Mn$_B$ pairs. 
As can be seen from Tables~\ref{tab:II}  and \ref{tab:III}, the Mn$_A^{2+}$ has
non-equivalent equilibrium positions in the cell where Mn$_B^{4+}$ is present. 
The superexchange interaction between Mn$_B^{4+}$ and Mn$_A^{2+}$  strongly depends 
on the displacement $d_A$.
Compared to the closer situation, the case with ions far from each other 
corresponds to a higher energy $\Delta E$.
As shown in detail in the following section, the 
 lattice polarization
increases the number $N_2(P_r,E)$ of Mn$_A^{2+}$ ions lying farther from Mn$_B^{4+}$ ions.  This
leads to a positive $\chi_{1,zzz}$. 
In the next section we show that this mechanism seems to be unimportant 
for the system studied in Ref.\cite{Shvartsman08}, but it may be very effective 
for a system where the interaction energy between electric Mn$_A^{2+}$-dipole
and $P_r$ is comparable with  $\Delta E$.

\section{Calculation of non-linear magneto-electric susceptibility}

Here we quantitatively consider the ideas outlined in the previous section. This
section is a little technical and can be safely skipped by someone not 
interested in the details of computation.

\subsection{The first mechanism} 
In order to estimate its contribution, we follow the ideas of Ref.\cite{Hou65}.
The Mn$_{B}^{4+}$ ion is much
more sensitive to the local strain than Mn$_{A}^{2+}$. In
an octahedral coordination it has a $^{4}A_{2}$ orbital singlet ground
state, and the effective spin 3/2 Hamiltonian has the form 
\begin{eqnarray}
\hat{H}_{B} & = & \mu_{B}g_{zz}H_{z}\hat{S}_{z}+\mu_{B}g_{\perp}
(H_{x}\hat{S}_{x}+H_{y}\hat{S}_{y})+\nonumber \\
 & + & D\left[\hat{S}_{z}^{2}-\frac{1}{3}S(S+1)\right]-
 \mu_{B}^{2}\sum_{\mu\nu}\Lambda_{\mu\nu}H_{\mu}H_{\nu}\ ,
\label{HB}
\end{eqnarray}
where 
$g_{\mu\nu}  
= 
g_{s}-2\lambda k\Lambda_{\mu\nu}$,  
$D=-\lambda^{2}\left(\Lambda_{zz}-\Lambda_{xx}\right)$, 
$\Lambda_{\mu\nu} = 
\sum_{n\neq0}\frac{\left\langle \psi_{0}\right|
\hat{L}^{\mu}\left|n\right\rangle \left\langle n\right|\hat{L}^{\nu}
\left|\psi_{0}\right\rangle }{E_{n}-E_{0}}$,
$g_{s}=2.0023$ is the spin gyromagnetic ratio, $\lambda$ the spin-orbit
coupling constant, $\left|\psi_{0}\right\rangle $($\left|n\right\rangle $)
is the ground(excited) state, $E_{0}$ and $E_{n}$ are the corresponding 
energies, $\hat{L}^{\mu}$ is the $\mu$-th component of the orbital
moment operator, $k$ is the orbital reduction factor. Using the ligand 
field theory\cite{kuzmin91,Kuzian06}, which perturbatively
takes into account $p-d$ hybridization between paramagnetic ion and
surrounding ligands, we may obtain\cite{Glinchuk07} 
\begin{eqnarray*}
\Lambda_{zz} & = & - 
\Delta g_{cub} 
\left(1-\kappa e_{33}\right)/2\lambda ,\\
\Lambda_{xx} & = & - 
\Delta g_{cub} 
\left(1- 
5\kappa e_{33}/2\right)/2\lambda ,
\end{eqnarray*}
 where $\Delta g_{cub}=g_{cub}-g_{s}$ refers to  the $g$ -value for\ undistorted
cubic lattice, $\kappa\sim-3.5,-4$ is the exponent of $p-d$ hopping
dependence on distance\cite{Harrison}. 
The dependence on electric field of the previous equations is only in 
the strain $e_{33}$. For multiglass samples we may
assume that both, the net polarization $P_r$ and the
electric field are directed along the axis $Z$, while $\mathbf{P}=\mathbf{P}_{r}+\left(\varepsilon-1\right)\mathbf{E}/4\pi$.
With an external magnetic field also applied along the $Z$-axis, we compute the partition
function $Z_B$ and the free energy density 
$F_{B}=-(y/v_{c})\theta\ln Z_{B}$, with  $y/v_{c}$ the density
of Mn$_{B}^{4+}$ ions, $v_{c}$ the host lattice
unit cell volume and $\theta\equiv k_{B}T$. From this free energy, we finally 
derive the linear contribution to the magnetic susceptibility (Eq.(\ref{eq:chiE})):
\begin{equation}
\chi_{1,zzz}\approx-\frac{y}{v_{c}}\mu_{B}^{2}\Lambda_{zz}^{\prime}
\left\{ \frac{g_{zz}\lambda}{\theta}\left[5 k+\frac{3g_{zz} \lambda}{2\theta}\right]-
2\right\},\label{eq:chi1app}
\end{equation}
where we have taken into account that $D/\theta\ll1$ for $T=10$ K.
The (double-)prime indicates the (second-) derivative with respect to $E$.
Similarly, for the second order contribution, we have 
\begin{eqnarray}
\chi_{2,zzzz} & \approx & \frac{y\mu_{B}^{2}}{v_{c}}
\left\{  k \left(\lambda\Lambda_{zz}^{\prime}\right)^{2}\left[10 k+
12g_{zz}\lambda/\theta \right]/\theta \right.\nonumber \\
 & - & \left. \Lambda_{zz}^{\prime\prime}\left[\frac{g_{zz}\lambda}{\theta}\left(5k+
 \frac{3g_{zz} \lambda}{2\theta} \right)-2\right]\right\} .
 \label{eq:chi2zzzz}
 \end{eqnarray}
When we substitute $\kappa=-3.5$, $g_{cub}=1.992$\cite{Laguta07},
$\lambda\approx135$ cm$^{-1}$,\cite{abragam}, $k\approx 1$, SrTiO$_{3}$ lattice
parameter $a=3.9$ \AA{}, ($v_{c}=a^{3}$), and $c/a\approx1.002$\cite{Kondakova09},
$\varepsilon(T=10\,\mathrm{K})\approx1500$, and a net polarization $P_{r}\approx0.7$, 
$\mu\mathrm{C/m^{2}}=2100 \ \mathrm{esu/cm^{2}}$\cite{Tkach05},
we obtain the values presented in Eqs. (\ref{eq:chi1val1}) and (\ref{eq:chi2val1}).

\subsection{The second mechanism} It involves a pair of Mn$_A$-Mn$_B$ 
ions. 
\begin{figure}[htbp]
\onefigure[scale=0.45]{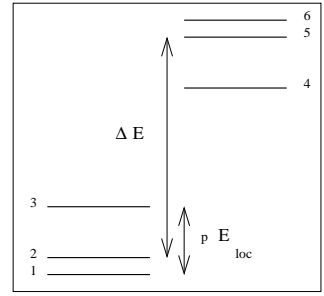} 
\caption{\label{Fig2} Changes of Mn$_A$-Mn$_B$ pair energies in polarized
medium. Here $E_{loc}=E+\frac{4\pi}{3}\gamma P $ is positive, however it can be  negative, depending on the
relative direction of polarization with respect to the pair Mn$_A$-Mn$_B$. The Lorentz
factor $\gamma$  accounts for the deviation
of the local field from the simple cubic case, $p\approx Z_{A}d_{A}$
is the dipole moment of Mn$_{A}^{2+}$, $Z_{A}\approx2$ is its dynamic
charge.}
\end{figure}
There are six available positions for Mn$_A$, three of them are closer to 
Mn$^{4+}_B$ and lower in energy: the number of Mn$^{2+}_A$ in deep (shallow)
wells is $N_{1(2)}$. Writing $N_1=n_1+n_2+n_3$, and $N_2=n_4+n_5+n_6$,
one has in absence of polarization $N_2/N_1=\exp(-\Delta E /\theta)$.
However with polarization, there is some lift of degeneracy, as can be seen on 
Fig. \ref{Fig2}, this leads to a redistribution of level occupancies, and 
$N_{1(2)}$ acquires a dependence on the net polarization $P_r$ and on the 
external electric field $E$.
With $N_1+N_2=xyz$, (
$z=8$ is the coordination number), the susceptibility reads
\begin{equation}
\chi = \chi_0 +\chi_1 xyz +\left(\chi_2-\chi_1\right) N_2(P_r,E),
\label{eq:chi2}\end{equation}
where $\chi_0$ comes from contribution of everything but Mn$_A$-Mn$_B$ pairs, 
while $\chi_{J}$($J=1,2$) is the susceptibility of a 
pair $S_{A}=5/2 - S_{B}=3/2$ coupled by the exchange interaction $J_{AB}(d_A)$, 
which depends on the Mn$_A$ position \cite{details}.
The dependence of $\chi$ on the electric field comes from 
$N_2(P_r,E)$. Defining $n(t)\equiv (2+\mathrm{e}^{-t})/(2+\mathrm{e}^{t})$, 
one has 
\begin{equation}
N_2(P_r,E)=\frac{xyz}{2} 
\left[ \frac{1}{n(t)\mathrm{e}^{\Delta E/\theta}+1} +
\frac{1}{n(-t)\mathrm{e}^{\Delta E/\theta}+1}  \right]
\label{eq:N2}\end{equation}
with $t=pE_{loc}/\theta $, 
$ p E_{loc} \approx c E +\Delta_r$, 
$c= p \varepsilon \gamma /3$ and $\Delta_r= 4 \pi \gamma p P_r /3$, 
($ \varepsilon \gg 1$ was used). See the caption of Fig.~\ref{Fig2} 
for definitions. Finally,
making a limited development up to second order in electric field $E$,  we obtain
\begin{eqnarray}
\chi_{1,zzz} & = & \left(\chi_{2}-\chi_{1}\right)N_2^{\prime}(P_r,0),
\label{eq:chi1zzz2}\\
\chi_{2,zzzz} & = & \left(\chi_{2}-\chi_{1}\right)N_2^{\prime\prime}(P_r,0).
\label{eq:chi2zzzz2}
\end{eqnarray}
Substituting the values from Table \ref{tab:II}: $\Delta E\approx-123+144.6=21.6$
meV, $J_{1}\approx1.7$~meV, $J_{2}\approx0.5$~meV,  and $\gamma \approx -0.2$\cite{VuGlin90},
this gives $\Delta_{r} \approx -14$~meV, and
a very tiny contribution of this second mechanism to the 
susceptibilities $\chi_1 \approx  7.1\cdot10^{-15}$, 
$\chi_2 \approx  1.6\cdot10^{-15}$. 
However, since $N_2^{\prime}$
is strongly dependent on $\Delta E - \Delta_{r}$, 
the second mechanism could be of the same order 
than the first one
or even exceed it for $\Delta E \approx \Delta_{r}$
(then $\chi_1 \approx  4.6\cdot10^{-7}$, 
$\chi_2 \approx  1.7\cdot10^{-7}$):
 this equality can be realized for 
other concentrations of manganese in SrTiO$_{3}$ host, or for other systems 
(e.g. K$_{1-x}$Mn$_{x}$Ta$_{1-y}$Mn$_{y}$O$_{3}$\cite{Kleemann09}).

\section{Summary}

We have compared the magnetic properties of two types of ceramic samples
of manganese doped SrTiO$_{3}$. Based on the data of ESR measurements
we conclude that the spin-glass behaviour is observed only in samples
containing an appreciable percentage of Mn$_{B}^{4+}$ ions substituting
for Ti, in addition to Mn$_{A}^{2+}$ substituting for Sr. Using LSDA+$U$
supercell calculation we have shown that the exchange interaction between
Mn$_{A}^{2+}$ impurities is an order of magnitude smaller than those
for Mn$_{B}^{4+}$-Mn$_{A}^{2+}$ and Mn$_{B}^{4+}$-Mn$_{B}^{4+}$pairs.
The analytic many-body calculations have shown that the reason for
this difference is the interference of various exchange paths for the
Mn$_{A}^{2+}$-Mn$_{A}^{2+}$
pairs combined with different geometry, Mn-O distances, and stability
of the Mn$_{A}^{2+}$ ground state configuration in comparison to other
pairs. We  conclude that the presence of Mn$_{B}^{4+}$ ions
is essential for the formation of a collective magnetic state at low
temperature. We propose two microscopic mechanisms of magnetoelectricity
in SrTiO$_{3}$:Mn which involve Mn$_{B}^{4+}$ ions.

\acknowledgments
The authors thank M.D. Glinchuk for fruitful discussions,
 the PICS program (Contracts CNRS No. 4767, NASU
No. 267) and grant MSMT CR (Project No. 1M06002) 
for financial support and the IFW Dresden (Germany) which allowed us 
to use their computer facilities. The institutional research
plan AVOZ10100521 is acknowledged.


\begin{thebibliography}{0}
\bibitem{details} We have
\begin{equation}
\chi_{J}=\sum_{S=\left|S_{A}-S_{B}\right|}^{\left|S_{A}+
S_{B}\right|}\frac{\left(2S+1\right)\exp\left(-E_{S}/\theta\right)}
{\sum_{S^{\prime}}\left(2S^{\prime}+
1\right)\exp\left(-E_{S^{\prime}}/\theta\right)} \chi_{S}
\end{equation}
with
$E_{S}=\frac{1}{2}J_{AB}[S(S+1)-S_{A}(S_{A}+1)-
S_{B}(S_{B}+1)]$,
the magnetic energy of the pair corresponding to the total spin $S$, 
$\chi_{S}=\left(\mu_{B}g_{S}\right)^{2}S\left(S+1\right)/(3v_{c}\theta)$
is the corresponding magnetic susceptibility, and the gyromagnetic ratio is
$g_{S}=\frac{1}{2}\left[g_{A}+g_{B}+\left(g_{A}-g_{B}\right)
\frac{S_{A}\left(S_{A}+1\right)-S_{B}\left(S_{B}+1\right)}{S\left(S+1\right)}\right]$.
\end{thebibliography}
\end{document}